\title{Assessing the Significance of Model Selection in Ecology}
\documentclass[12pt]{paper}
\usepackage{graphicx}
\usepackage{float}
\usepackage{url}
\usepackage{dirtytalk}
\usepackage{bm}
\usepackage{color}
\usepackage{xcolor}
\usepackage{rotating}
\usepackage{breakcites}
\usepackage{natbib}
\usepackage{tikz}
\usepackage{authblk}
\author{Edward Wheatcroft\thanks{corresponding author: e.d.wheatcroft@lse.ac.uk}}
\affil{\small Centre for the Analysis of Time Series, London School of Economics and Political Science}
\date{\today}

\begin{document} 
\sloppy
\maketitle
\begin{abstract}
Model Selection is a key part of many ecological studies, with Akaike's Information Criterion the most commonly used technique.  Typically, a number of candidate models are defined a priori and ranked according to their expected out-of-sample performance.  Model selection, however, only assesses the relative performance of the models and, as pointed out in a recent paper, a large proportion of ecology papers that use model selection do not assess the absolute fit of the `best' model.  In this paper, it is argued that assessing the absolute fit of the `best' model alone does not go far enough.  This is because a model that appears to perform well under model selection is also likely to appear to perform well under measures of absolute fit, even when there is no predictive value.  A model selection permutation test is proposed that assesses the probability that the model selection statistic of the `best' model could have occurred by chance alone, whilst taking account of dependencies between the models.  It is argued that this test should always be performed before formal model selection takes place.  The test is demonstrated on two real population modelling examples of ibex in northern Italy and wild reindeer in Norway.
\end{abstract}


\section{Introduction}
Model selection forms a key part of a large proportion of publications in ecology journals.  This is particularly true in population modelling studies in which generalised linear models (GLMs) are typically tested with different combinations of potential predictor variables (\cite{thieme2018mathematics,ECY:ECY20048561598,imperio2013climate}).  In a large number of cases, Akaike's information criterion (AIC), or its adjusted version for small samples AICc, is used to compare the relative performance of different combinations of variables (henceforth `models').  The model with the most support according to the information criterion is then usually selected for further use or as a conclusion in itself. There is a noted tendency, however, to neglect to test whether \emph{any} of the models are indeed useful or even `significant'.  After all, the best of a bad bunch of models is still a bad model.  It was found by \cite{JPE:JPE13060} that, out of 119 ecology papers considered that use information criteria to compare the performance of different models, only 55 included some measure of the absolute goodness of fit. The authors of that paper suggest both that some measure of absolute performance should be shown and that the null model (i.e. a statistical model with no explanatory variables) should always be included as a benchmark with which to compare the performance of each of the candidate models.  This idea is expanded upon in \cite{wheatcroft_null_model} in which more flexible benchmark models are suggested as alternatives to the null model. \par

Whilst it is essential that some measure of the absolute goodness of fit of the `best' model is included, it is argued here that doing so does not solve the problem entirely, due to implicit multiple testing that is not taken into account.  Suppose that a statistical test measures the significance of the `best' model, which has been determined by an information criterion or some other method of model selection.  Whilst, in this case, only one formal test is actually performed, the model of interest has already been determined as one that appears to perform relatively well under model selection. Since there is a close relationship between model selection techniques such as information criteria and formal statistical significance tests, those that perform well under the former tend also to perform well under the latter. Crucially, this is true both when the model is actually informative. i.e. would perform better than the null model out-of-sample, and when it appears to be informative only by random chance. Therefore, in the latter case, the probability that a statistical test on the `best' model is wrongly found to be significant is inflated.  In statistical terminology, this means that the probability of a type I error is increased. It is argued in this paper that multiple testing needs to be accounted for when assessing the significance of each of the models and a framework is introduced with which to do this. \par

The distinction between assessing the relative and absolute values of a set of candidate models is well known.  For example, in the context of ecology, it was pointed out by \cite{symonds2011brief} that, in model selection, `you can have a set of essentially meaningless variables and yet the analysis will still produce a best model'.  They therefore suggest that it is `important to assess the goodness of fit ($\chi^{2}$, $R^{2}$) of the model that includes all the predictors under study, arguing that 'if this global model is a good fit, then you can rest assured that the best approximating model will be a good fit also'. This approach seems somewhat ad-hoc since a global model with a large enough number of parameters will always appear to provide a good fit in-sample, regardless of how informative each of the variables are.  As an explanation of why testing the significance of the `best' model is often neglected, \cite{burnham2004multimodel} suggest that, historically, it has often been assumed that there is a single `true' model and that that model is in the candidate set. The Bayesian derivation of the Bayesian Information Criterion (BIC), for example, works under this assumption (\cite{burnham2004multimodel}). If the assumption holds, with enough data, one can eventually expect to select the true model, and that model will, by definition, provide a good fit.  In practice, few people believe that the `true' model is ever likely to be a member of the candidate set. Another suggested approach is to ensure that each variable included in the model is carefully justified \emph{a priori}, such that only variables with a high chance of being informative are included (\cite{burnham2001kullback,burnham2004multimodel}).  Whilst this is a sensible suggestion, it does not solve the problem since, whilst those variables may be expected to be important, \emph{a priori}, this may not be reflected in the models once the data have been considered. \par

The statistical literature on multiple testing is considerable.  Perhaps the most well known approach to the problem is the Bonferroni correction which makes a simple adjustment to the significance level according to the number of hypotheses that are tested (\cite{bonferroni1936teoria}).  Other methodologies, such as the Bonferroni-Holm method \cite{holm1979simple} and Benjamini–Hochberg and Sidak corrections (\cite{benjamini1995controlling,vsidak1967rectangular}), for example, control the order in which tests are applied to limit the number of overall tests, producing a uniformly more powerful approach. These are discussed further in section~\ref{section:multiple_testing}.

A weakness of the above approaches is that they assume that each of the hypotheses are independent of each other.  If there is dependency between a set of hypotheses, the probability of committing a type I error in at least one of those hypotheses does not generally grow as quickly as when they are independent.  Such methods are therefore too conservative in such cases, with the result that the probability of rejecting an informative model is increased, i.e. a type II error is committed (\cite{nakagawa2004farewell}). To attempt to overcome this problem, the Westfall-Young procedure uses permutation tests to adjust the p-values in multiple correlated hypothesis tests, whilst taking account of the dependency between the hypotheses \cite{westfall1993adjusting}. This provides a test which is far more powerful in such cases. \par

In ecology, it is common to define candidate models as different combinations of the same set of candidate variables in a generalised linear model (\cite{bolker2009generalized}). There is therefore a strong degree of dependency between the candidate models and so the Bonferroni correction is unsuitable (along with other similar procedures).  In this paper, two permutation tests are proposed, which are based on the Westfall-Young procedure.  The first test, named the \emph{single model permutation test}, assesses the significance of individual models on the basis of a model selection statistic.  This is then extended to define another test, called the \emph{model selection permutation test}, that measures the significance of the entire model selection procedure, whilst taking into account the dependencies between the candidate models.  The result of the first test is an individual p-value for each model whilst the result of the second test is a single p-value relating to the model selection procedure itself.  The idea is then that, if the p-value of the model selection permutation test is smaller than the chosen significance level, the whole model selection procedure can be considered to be `significant', that is the probability of finding a model selection statistic as good or better than that of the `best' model by chance is small.  Model selection can then go ahead with the reassurance that the performance of the `best' models is unlikely to have occurred simply due to random chance. \par

The model selection permutation test proposed in this paper has been utilised in another recent paper entitled `Effects of weather and hunting on wild reindeer population dynamics in Hardangervidda National Park' on which the author of this paper is also named.  In that paper, the test is referred to as a `sanity check' test and a reference to this paper is provided.  As such, some of the analysis from that paper is reproduced here with the primary focus here being the application of the proposed tests. Additionally, in this paper, an example is used in the form of a population modelling analysis of ibex populations in the northern part of Italy which was taken from an existing paper published in 2004 (\cite{ECY:ECY20048561598}). \par

\section{Methods}

\subsection{Forecast Evaluation and Model Selection}
It was noted by \cite{JPE:JPE13060} that authors commonly neglect to include a measure of the absolute performance of the `best' model alongside a model selection procedure.  Of those papers that do include such a measure, the vast majority were found to use $R^{2}$, adjusted $R^{2}$ or related measures. Although the low number of cases in which no absolute measure of fit is provided is concerning, those measures that \emph{are} commonly used for this purpose can be problematic themselves. AIC and its corrected version are founded in information theory and approximate the expected difference in information loss from approximating the underlying system with different candidate models. Since `information' in this case relates to the probability or probability density assigned to the outcome, it is therefore a measure of probabilistic performance. It can be noted that generalised linear models, as commonly used in ecology, naturally provide a set of probabilistic forecasts. $R^{2}$ and similar related metrics, however, are measures of deterministic performance, i.e. they only consider a forecast to be a single number.  This means that, whilst models are selected according to the performance of the resulting \emph{probabilistic} forecasts, they are evaluated as \emph{point} forecasts.  This seems like an inconsistent approach to the forecasting problem as a whole. \par

In fact, probabilistic forecasts can contain a great deal of information that cannot be communicated in point forecasts.  In the case of a Gaussian forecast distribution, for example, the variance can be of great value in understanding the uncertainty in the point estimate defined by the mean.  For more complex forecast distributions, a single number such as the mean may be entirely inadequate.  Consider, for example, a herd of terrestrial animals that, according to a probabilistic forecast distribution, is equally likely to be on either side of a large lake that runs from north to south.  It is difficult, in this case, to define a single number from the distribution that represents a useful point forecast.  After all, it makes little sense to predict the mean of that distribution since it would fall within the lake, an area in which there is little or no chance of the herd residing. Equally, it would make little sense to forecast that the herd will be on a particular side of the lake since each are deemed equally likely.  In summary, to use a point rather than a probabilistic forecast, information must be discarded. \par

In addition to the issues described above, measures of the predictive performance of point forecasts tend to be fraught with problems.  For example $R^{2}$, which appears to be the most commonly used measure in ecology papers, is widely known to be a poor measure of forecast performance (\cite{wheatcroft2015improving}). Firstly, the correlation is insensitive to scale. This means that, if two variables are correlated, it doesn't necessarily mean that one is a good predictor of the other. For example, a set of temperature forecasts measured in Fahrenheit when the outcomes are measured in Celsius may still have a high $R^{2}$ value.  This has been widely acknowledged and, for example, Murphy describes $R^{2}$ as a measure of \emph{potential} rather than absolute skill (\cite{murphy1989skill}).  Secondly, other well known problems with using correlation coefficients apply. For example, influential observations can greatly increase the correlation between two variables without much, or any, actual improvement in predictive performance (\cite{wheatcroft2015improving}). \par

\subsection{Evaluating Probabilistic Forecasts} \label{section:evaluation}
Probabilistic forecasts are usually evaluated using functions of the forecast and the outcome called \emph{scoring rules}.  A wide range of scoring rules have been proposed and there is still some debate over which are the most appropriate (\cite{gneiting2007strictly}).  A property of scoring rules generally considered to be of high importance is called \emph{propriety}.  A score is \emph{proper} if it is optimised in expectation when the distribution from which the outcome is drawn is issued as the forecast (\cite{Importance_proper}).  Propriety would therefore discourage a forecaster in possession of that forecast distribution from issuing a different one to achieve a better score.  It is worth noting that no similar property exists for measures of the performance of point forecasts.  For example, $R^{2}$ needn't favour a forecast based on the true distribution of the outcome. \par

An example of a proper scoring rule is the \emph{ignorance score} (\cite{Good:1952,Roulston02evaluatingprobabilistic}) defined by 
\begin{equation}
\mathrm{IGN}=-\log_{2}(p(Y))
\end{equation}
where $p(Y)$ is the probability density placed on the outcome.  The ignorance score is negatively oriented and hence smaller values indicate better forecast skill. The score is also \emph{local} because it only takes the probability at the outcome into consideration (\cite{gneiting2007strictly}) and, in fact, can be shown to be the only scoring rule that is both proper and local (\cite{bernardo1979expected}). An advantage of the ignorance score is in its interpretation.  The difference in the mean ignorance between two sets of forecasts can be interpreted as the base 2 logarithm of the ratio of the density placed on the outcome by each, measured in bits.  For example, if the mean ignorance of one set of forecasts is 3 bits smaller than another, it places $2^{3}$ times more probability density on the outcome, on average. The ignorance score is used in the results section of this paper alongside leave-one-out cross-validation. \par 

\subsection{Approaches to Model Selection}
Model selection is a key part of many studies in a wide range of disciplines, including ecology (\cite{johnson2004model}).  The standard approach is to define a set of candidate models \emph{a priori} and to attempt to rank them according to how well they would generalise out-of-sample. The basis of model selection techniques is founded on the observation that a fair comparison is needed between models with different numbers of parameters. If an extra parameter is added, the fit of the model will necessarily improve in-sample but may be `overfitted' and will not improve out-of-sample. Model selection techniques therefore attempt to account for this issue. \par

Model selection techniques typically fall into two different categories.  Information criteria weigh up the in-sample fit of the model with the number of parameters to be selected such that extra parameters are penalised.  Cross-validation, on the other hand, divides the dataset such that parameter selection is always performed on data that are distinct from those on which the performance of the model is tested. \par

By far the most commonly used information criterion in ecology is Akaike's Information Criterion (AIC) and its corrected version for small samples AICc (\cite{AIC,AIC_model_selection}).  AIC is given by
\begin{equation}
\mathrm{AIC}=-2\log(\hat{L})+2K
\end{equation}
where $\hat{L}$ is the maximised likelihood and $K$ is the number of parameters selected.  In each case, the model with the lowest AIC is considered to be the most appropriate when applied out-of-sample.  For small sample sizes, however, AIC is slightly biased and thus a corrected, unbiased, version is often used.  The corrected version $\mathrm{AIC_{c}}$ (\cite{claeskens_hjort_2008}) is defined by
\begin{equation}
\mathrm{AICc}=-2\log(\hat{L})+\frac{2K(K+1)}{n-K-1}.
\end{equation}

Cross-validation takes a different approach to model selection. Here, the data are divided into two sets: a training set, over which the parameters are selected, and a test set, on which the model is tested with those parameters.  The process is then repeated with different subsets of the data set used as the training and test sets.  In leave-one-out cross-validation, the test set consists of a single point and the training set consists of each of the other points.  This process is repeated such that each data point forms the test set exactly once. Leave-one-out cross-validation can be used alongside any method of forecast evaluation and, in this paper, is performed with the ignorance score such that the forecasts can be evaluated probabilistically. In fact, this approach can be shown to be asymptotically equivalent to AIC but will usually be expected to give a different ordering of models for finite sample sizes (\cite{stone1974cross}). \par  

This paper suggests a two step process to model selection.  First, the significance of the model selection procedure should be assessed at some pre-defined level to assess the probability that a statistic at least as favourable than that of the `best' model could have occurred by chance, given the candidate models.  If the model selection procedure is found to be significant, normal model selection should then take place and the best model(s) chosen.  By taking the first step, confidence can be had that the information contained in the models is indeed informative. \par

It is important to note that, even if the model selection procedure is found to be significant, it is not necessarily the case that any of the model(s) are fit for their required purpose (e.g. population management).  To determine whether the models are fit for purpose would require further analysis and consideration beyond the scope of this paper. \par

\subsection{Permutation Tests}
A permutation test is a nonparametric statistical test in which the significance of a test statistic is obtained by calculating its distribution under all different permutations of the set of observed outcomes.  For example, a permutation test for the slope parameter of a simple linear regression would be performed by permuting the positions of the $y$ values (the dependent variable), keeping the $x$ values (the predictor variables) in their original positions and calculating the slope parameter under all possible combinations of $y$.  The position of the slope parameter that has been calculated from the data in their original positions would then be compared to this distribution to calculate a p-value.  In practice, it is often computationally prohibitive to consider all possible permutations and thus permutations are randomly chosen a fixed number of times.  Such tests are called \emph{randomised permutation tests}.  Permutation tests have a number of advantages over standard parametric tests. Unlike the latter, no assumptions about the distribution of the test statistic under the null hypothesis are required since the method draws from the exact distribution.  Permutation tests thus give an exact test and, as such, randomised permutation tests are asymptotically exact. The general nature of permutation tests allows them to be applied in a wide range of settings without knowing the underlying sampling distribution.  In this paper, two types of permutation test are demonstrated.  The first assesses the significance of a single model without taking into consideration the other models in the model selection procedure whilst the second assesses the significance of the entire model selection procedure and thus takes into account multiple testing. \par

\subsection{Multiple Testing} \label{section:multiple_testing}
The problem of multiple testing is well-known and has been widely studied.  Remedies to the problem typically involve adjustments to the p-value of each hypothesis to reflect the number that are tested.  Much of the literature on multiple testing focuses on controlling the \emph{familywise error rate} (FWER) $\alpha_{f}$, defined as the probability of wrongly rejecting at least one of the hypotheses.  Whilst, under standard hypothesis testing, $\alpha_{f}$ usually grows with the number of hypotheses, the aim here is usually to limit the FWER to $\alpha_{f}$.  Perhaps the most common approach to the problem is the Bonferroni correction which adjusts the required significance level for each test to $\frac{\alpha_{f}}{m}$ where $m$ is the number of hypotheses tested.  A major weakness of the Bonferroni correction, however, is that it assumes that each of the significance tests are independent of each other.  When this is not the case, the test is too conservative and the true FWER is less than $\alpha_{f}$, resulting in a loss of power. Several modifications to the Bonferroni correction, such as the Bonferroni-Holm (\cite{holm1979simple}), Benjamini-Hochberg (\cite{benjamini1995controlling}) and Sidak (\cite{vsidak1967rectangular}) corrections, have been proposed that aim to increase the power by adjusting the order in which hypotheses are considered.  None of these approaches take into account dependency between hypotheses, however. \par

An alternative approach to multiple testing was proposed by Westfall and Young in 1993 and aims to account for dependency between tests (\cite{westfall1993adjusting}).  The approach makes use of permutation tests by randomly permuting the outcomes and calculating adjusted p-values for each hypothesis.  An adjusted p-value for the $ith$ hypothesis is given by 
\begin{equation}
\tilde{p}_{i}=Pr(\min_{1\leq j \leq m} P_{j}<p_{i}|H_{0}^{C})
\end{equation}
where $p_{i}$ denotes the observed p-value for the $ith$ test, $H_{0}^{C}$ is the `complete' null hypothesis that all null hypotheses are true and $P_{j}$ is the p-value of the $jth$ hypothesis under a given permutation of the outcomes.  The adjusted p-value of the $ith$ hypothesis corresponds to the probability of obtaining a p-value as small or smaller from at least one of the $m$ hypotheses that are tested simultaneously. \par

\subsection{A Single-model Permutation Test} \label{section:single_model_permutation_test}
A permutation test is now described with which to test the significance of individual models in a model selection procedure. The test is performed by comparing the observed model selection statistic with the distribution of that statistic under the null hypothesis that the outcomes are independent of the model predictions.  An approximate p-value is calculated by counting the proportion of permutations in which the model selection statistic is smaller (assuming a negatively oriented statistic) than the observed statistic. The single-model permutation test is formally described below: \par

\begin{enumerate}
\item Calculate the model selection statistic $M$ under the original ordering of the outcomes. 
\item Set $j=1$
\item Randomly permute the outcomes, ensuring that none of them fall into their original positions.
\item Calculate the model selection statistic $\tilde{M}_{j}$ under the new ordering.
\item Set $j=j+1$
\item Repeat steps two to four until $j=J$.
\item Calculate the approximate p-value $p=\frac{1}{J} \sum_{j=1}^{J} I(\tilde{M}_{j}<M)$, where $I$ is the indicator function.
\end{enumerate}

In fact, the test outlined above can be considered a standard permutation test and, as such, is not particularly novel and is somewhat similar to that of the Westfall-Young permutation test.  However, whilst that test uses individual p-values to calculate adjusted p-values for each hypothesis (or model), the above test uses model selection statistics which do not necessarily naturally have p-values associated with them.  

The single model permutation test provides a simple basis with which to assess the significance of a single model. Note that, for a single model, when the chosen model selection statistic is an information criterion, the penalty for the number of parameters is always constant and therefore the test is equivalent to performing a permutation test on the log-likelihood.  However, in the next section, the test is extended to multiple models with different numbers of parameters and it is here in which the value of permutation tests for model selection statistics becomes apparent. \par

\subsection{A Model-selection Permutation Test} \label{section:sense_check}
A permutation test for an entire model selection procedure is now defined. The aim here is to estimate the probability that the `best' model selection statistic could have occurred by chance.  We call this test the \emph{model-selection permutation test}. \par

Under the model selection permutation test, the outcomes are randomly permuted as they are for the single model permutation test.  Here, the null hypothesis is that the outcomes and the model predictions are independent for all tested models.  For a given permutation of outcomes, a model selection statistic is calculated for each model. The comparison of interest is between the observed `best' model selection statistic and the statistic of the `best' model under each permutation. The p-value is estimated by counting the proportion of permutations in which the model selection of the `best' model is more favourable than that of the `best' model under the true ordering of the outcomes.  Formally, the procedure is performed as follows:

\begin{enumerate}
\item Calculate the model selection statistic for each model $M_{1},..,M_{m}$ under the original ordering of the outcomes. 
\item Set $j=1$
\item Randomly permute the outcomes, ensuring that none of them fall into their original positions.
\item Calculate the model selection statistic for each model $\tilde{M}_{j,1},..,\tilde{M}_{j,m}$.
\item Set $j=j+1$
\item Repeat steps two to four until $j=J$.
\item Calculate an estimated p-value $p=\frac{1}{J} \sum_{j=1}^{J} I(\mathrm{min}(\tilde{M}_{j,i},..,\tilde{M}_{j,m})< \mathrm{min}(M_{1},..,M_{j}))$ where $I$ is the indicator function.
\end{enumerate}

\subsection{Experiment One: Demonstration of Type I Error Inflation}
The aim of experiment one is to demonstrate that, in a case in which all models are, by construction, uninformative, the probability that the `best' model is `significant' increases with the number of candidate models. This represents, by definition, inflation in the probability of a type I error.  It is then demonstrated that, for the model selection permutation test, the probability of a type I error is consistent with the prescribed significance level and is not affected by the number of candidate models.  This is demonstrated in two cases: one in which the models are defined to be independent of each other and another in which there is dependency between models resulting from shared predictor variables. \par
 
The experiment is conducted as follows:  Let ${\bm y}=y_{1},...,y_{20}$ be a set of outcomes, each of which are independent, identically distributed draws from a standard Gaussian distribution $N(0,1)$. Let ${\bm X}=X_{1,m},..,X_{20,m}$ be the mth predictor variable of ${\bm y}$ which is also $iid$ standard Gaussian and is independent of ${\bm y}$.  Define a model to be some combination of predictor variables in a multiple linear regression with  ${\bm y}$ as the dependent variable.  As such, none of the models have any predictive value out-of-sample and thus the null model is, by design, the optimal choice.  AICc is calculated along with a p-value from the single model permutation test. The `best' model, according to AICc, is then defined to be significant if its p-value is less than $0.05$, i.e. it is significant at the 5 percent level. In addition, the model selection permutation test is performed at the 5 percent level. This procedure is repeated $256$ times and the proportion of repeats in which the `best' model is found to be significant and the proportion in which the model selection permutation test is found to be significant is calculated.  \par 

Two different cases are considered:
\begin{enumerate}
\item independent models - each model is a linear regression with one of ${\bm X_{1}},...{\bm X_{n}}$ as a single predictor variable. There are thus $n$ candidate models.
\item dependent models - $k$ candidate variables ${\bm X_{1}},...{\bm X_{k}}$ are defined and $n$ distinct combinations are randomly selected, without replacement, as candidate models.
\end{enumerate}
In the former case, by construction, each model is independent.  In the latter, however, since different candidate models have shared predictor variables, there is a dependency structure between models.  For each value of $k$, there are $2^{k}-1$ possible combinations of variables (excluding the null model) and thus only values of $n$ up to this value can be considered.  Therefore, for $n=2^{k}-1$, all of the possible combinations of variables are tested and only a subset are tested for $n \le 2^{k}-1$. \par

\subsection{Population Modelling Examples}
Two real population modelling examples from ecology are used to demonstrate both the single model and model selection permutation tests.  Both examples are published in existing papers and are presented here with the minimal details required to effectively demonstrate the methodology presented in this paper.  Further details can be found in the papers themselves.

\subsubsection{Experiment Two: Ibex}
The first population modelling example was published in \emph{Ecology} in 2004 in `Climate forcing and density dependence in a mountain ungulate population' (\cite{ECY:ECY20048561598}).  In that paper, the authors fit 20 different population models to attempt to explain changes in the ibex population of Gran Paradiso National Park in Northwestern Italy between the years of 1956 and 2000, using combinations of the following predictor variables:
\begin{itemize}
\item Current population.
\item Snow cover.
\item Interaction between snow cover and current population.
\end{itemize}
Ten different combinations of the three variables were fitted with both the Modified Stochastic Ricker and Modified Stochastic Gompertz models (defined in the appendix) such that a total of twenty models were assessed.  The \emph{relative population change} in year $i$ is defined as $R_{i}=\log(\frac{n_{i+1}}{n_{i}})$ where $n_{i}$ and $n_{i+1}$ are the population counts in years $i$ and $i+1$ respectively. The Modified Stochastic Ricker and Modified Stochastic Gompertz models are generalised linear models such that the relative population change is modelled as a linear function of the chosen predictor variables.  The Stochastic Ricker and Gompertz models differ only in the way they treat the current population size as a predictor variable. \par

AIC was calculated for each model based on its performance in predicting the relative population change (rather than the actual population size).   Although, in that paper, AIC was the only model selection statistic considered, here, for illustration, the models are also compared using leave-one-out-cross-validation with the mean ignorance score as the evaluation method (see section~\ref{section:evaluation}). The model selection statistics for each model are presented relative to that of the null model, i.e. with the statistic of the null model subtracted, such that a negative value indicates more support for the model than for the null model. \par

To demonstrate the two tests defined in this paper, the single model permutation test is performed for each model and an estimated p-value is calculated.  In addition, results from the model selection permutation test are shown to assess the credibility of the overall model selection procedure. \par

\subsection{Experiment three: Reindeer} \label{section:methods_reindeer}
This example comes from a study of the population of wild reindeer in Hardangervidda National Park in Southern Norway (\cite{Hardangervidda}).  The aim of the study was to attempt to understand the factors that cause the population to change over time. This was done using a Modified Stochastic Ricker population model (defined in the appendix) with various combinations of factors as inputs.  The following climatic factors were considered as potential predictors of the population:
\begin{itemize}
\item (a) Mean temperature over January and February.
\item (b) Days in February/March in which the temperature exceed $0\circ C$.
\item (c) The number of summer growing degree days from June to September (days above 5 degrees Celsius).
\item (d) The current size of the population (density dependence).
\item (e) The proportion of the population hunted and killed.
\item (f) Interaction between proportion killed and chosen weather variable.
\item (g) Interaction between population size and chosen weather variable.
\end{itemize}
The winter of 2010 was significantly colder than each of the other years in the data set and was found to be an influential observation (according to Cook's distance). Given this, the analysis was performed twice: with and without that year included. The corrected version of Akaike's Information Criterion (AICc) was used to rank the performance of the models.  \par

In this paper, the analysis is repeated and, for the purposes of demonstration, the models are also compared using the cross-validated mean ignorance score, as an alternative model selection technique. The analysis is performed with the year 2010 removed (see above).  Following the original paper, a slightly different approach is taken to that of the ibex example. Whilst, in the ibex case, the performance of the models was assessed in terms of prediction of the relative population change, in this case, forecasts of the actual population counts were produced. To do this, Monte-Carlo simulation was used with a large sample and forecast distributions were produced using kernel density estimation. \par

Both the single model permutation test and the model selection permutation test are performed in the context of both the cross-validated mean ignorance and the AICc for the original set of candidate models.  The experiment is then repeated with a subset of the models to demonstrate a case in which the model selection permutation test is not significant.  \par

\section{Results}

\subsection{Experiment One: Demonstration of Type I Error Inflation}
\begin{figure}[!htb]
    \centering
    \includegraphics[scale=0.44]{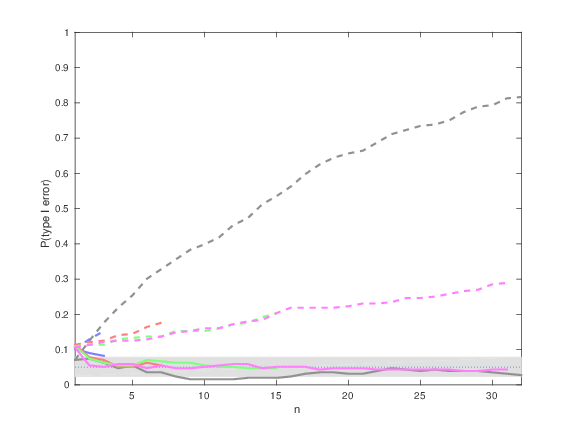}
    \caption{Proportion of repeats in which (i) the `best' model, as chosen by AICc, is significant at the 5 percent level under the single model permutation test (dashed lines) and (ii) the model selection permutation test (solid lines) is significant at the 5 percent level. Black lines show the results for the independent case and the blue red, green and magenta lines show the dependent case for $k$ equal to 1,2,3 and 4 respectively.}
    \label{fig:Random_model_selection_typeI_error}
\end{figure}

The results of experiment one are now presented. In figure~\ref{fig:Random_model_selection_typeI_error}, the dashed lines show, as a function of $n$, the proportion of repeats in which the `best' model, as selected by AICc, is found to be significant under the single model permutation test and the solid lines show the proportion of repeats in which the model selection permutation test is found to be significant.  Both tests are performed at the 5 percent level.  The grey area denotes the interval in which the proportions would fall with 95 percent probability if the underlying probability of a significant result were truly 5 percent.  If the proportion falls outside of this range, there is significant evidence that the probability of rejecting, and therefore committing a type I error, is different to the prescribed significance level. \par

As expected, as the number of candidate models is increased, the probability of a significant result for the `best' model is inflated beyond the prescribed significance level.  This is true of both the independent and dependent models cases.  In the latter case, the probability increases less quickly because fewer predictor variables are considered and therefore the probability of finding one that happens to be `significantly' correlated with the outcomes is reduced. This shows the importance of taking dependency between models into account.  The proportion of cases in which a significant result is found for the model selection permutation test is demonstrated to be consistent with the significance level of 5 percent. \par

\subsection{Experiment Two: Ibex}
The results of the model selection procedure for the ibex example are shown in table~\ref{table:AIC_rel_clim_ibex}.  Consistent with the original paper, columns headed by $b$, $c$ and $e$ indicate whether density dependence, snow cover and the interaction between the two, respectively, have been included in the model.  It is found that almost all of the models outperform the null model under both model selection methods.  Estimated p-values calculated using the single model permutation test are shown for each model based on the two model selection techniques considered.  In all cases, the estimated p-values are found to be extremely small.  \par

\begin{figure}[!htb]
    \centering
    \includegraphics[scale=0.44]{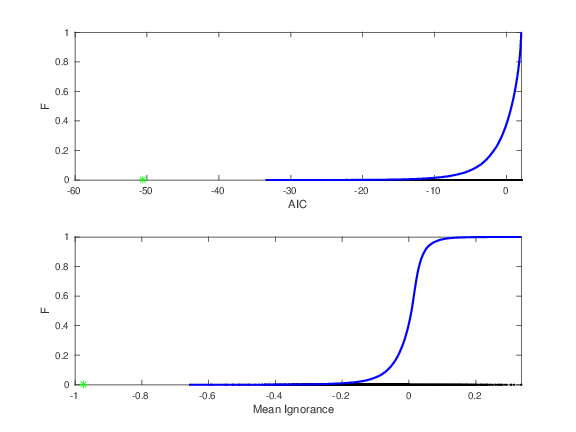}
    \caption{Top: Smallest AIC values from resampled data (black dots), their CDF and the smallest AIC from the observed data (green star) for the ibex example. Bottom: the same for cross-validated mean ignorance}
    \label{fig:ECDF_AIC_GP}
\end{figure}

Although it seems unlikely that the performance of the models could be explained simply through random chance, it is rigorous to use the model selection permutation test to assess the overall significance of the model selection procedure.  For both model selection techniques, out of $2^{16}=16384$ permutations tested, none were found in which the `best' model outperformed that for the observed outcomes and thus the estimated p-value is zero.  A CDF of the AIC of the `best' model (relative to the null model) under each permutation is shown in the top panel of figure~\ref{fig:ECDF_AIC_GP} along with the minimum AIC from the observed data set.  The equivalent, but with the cross-validated mean ignorance, is shown in the lower panel.  From, these results, it is clear that it is extremely unlikely that the `best model' in the model selection procedure occurred purely by chance.  Given its strong significance, confidence can be had that the results indicate genuine predictive skill. \par

\begin{sidewaystable}[]
\centering
\begin{tabular}{|l|ccccc|cc|cc|}
\hline
Model & b & c & e & Pars & DD  & AIC & p-value & Mean Ign & p-value \\
\hline
M11   & * & * & * & 7 & R & $-51.2$ & $0/(2^{16})$ & $-0.98$ & $0/(2^{16})$ \\
M12   & * & * & * & 7 & G & $-50.6$ & $0/(2^{16})$  & $-0.94$ & $0/(2^{16})$ \\
M13   &   & * & * & 5 & R & $-46.1$ & $0/(2^{16})$  & $-0.84$ & $0/(2^{16})$ \\
M14   &   & * & * & 5 & G & $-45.8$ & $0/(2^{16})$  & $-0.82$ & $0/(2^{16})$ \\
M15   & * &   & * & 5 & R & $-37.2$ & $0/(2^{16})$  & $-0.66$ & $0/(2^{16})$ \\
M4    &   & * & * & 3 & G & $-36.8$ & $0/(2^{16})$  & $-0.67$ & $0/(2^{16})$ \\
M2    & * & * & * & 4 & G & $-34.8$ & $0/(2^{16})$  & $-0.62$ & $0/(2^{16})$ \\
M16   & * &   & * & 5 & G & $-34.7$ & $0/(2^{16})$  & $-0.58$ & $0/(2^{16})$ \\
M3    &   & * & * & 3 & R & $-33.9$ & $0/(2^{16})$  & $-0.57$ & $0/(2^{16})$ \\
M18   & * & * &   & 5 & G & $-32.4$ & $1/(2^{16})$  & $-0.52$ & $1/(2^{16})$ \\
M1    & * & * & * & 4 & R & $-31.9$ & $0/(2^{16})$  & $-0.50$ & $1/(2^{16})$ \\
M5    & * &   & * & 3 & R & $-30.9$ & $0/(2^{16})$  & $-0.56$ & $0/(2^{16})$ \\
M6    & * &   & * & 3 & G & $-30.6$ & $0/(2^{16})$  & $-0.55$ & $0/(2^{16})$ \\
M8    & * & * &   & 3 & G & $-30.0$ & $0/(2^{16})$  & $-0.54$ & $0/(2^{16})$ \\
M7    & * & * &   & 3 & R & $-27.1$ & $0/(2^{16})$  & $-0.47$ & $0/(2^{16})$ \\
M17   & * & * &   & 5 & R & $-24.8$ & $0/(2^{16})$  & $-0.27$ & $44/(2^{16})$ \\
M9    &   &   & * & 2 & R & $-22.1$ & $0/(2^{16})$  & $-0.43$ & $0/(2^{16})$ \\
M19   &   &   & * & 3 & R & $-20.8$ & $0/(2^{16})$  & $-0.37$ & $0/(2^{16})$ \\
M10   &   &   & * & 2 & G & $-10.4$ & $34/(2^{16})$ & $-0.16$ & $368/(2^{16})$ \\
M20   &   &   & * & 3 & G & $-10.1$ & $85/(2^{16})$ & $-0.12$ & $445/(2^{16})$ \\
M0    &   &   &   & 1 &   & $0$     & -             & $0$     & - \\
\hline
\end{tabular}
\caption{AIC and cross-validated mean ignorance scores expressed relative to the null model for each ibex model.  Columns headed by $b$, $c$ and $e$ indicate whether density dependence, snow cover and the interaction between the two have been included in the model respectively.  Also shown are the total number of free parameters and whether the Ricker or Gompertz model has been used (R or G). The p-value of each model from the single model permutation test when using the AIC and the cross-validated mean ignorance are shown.  The models are listed in order of their AIC values.}
\label{table:AIC_rel_clim_ibex}
\end{sidewaystable}

\subsection{Experiment three: Reindeer}
The results of the model selection procedure for the reindeer case are shown in table~\ref{table:AIC_rel_clim_reindeer}. Here, those variables that are included in the model are indicated with a star.  The letters correspond to the variables listed in section~\ref{section:methods_reindeer}.  The AICc and mean ignorance (both shown relative to that of the null model) are shown for each model along with estimated p-values obtained from the single variable permutation test. \par
\begin{sidewaystable}[]
\fontsize{12}{12}\selectfont
\centering
\begin{tabular}{|l|cccccccc|cc|cc|}
%\multicolumn{1}{|c|}{} & \multicolumn{6}{|c|}{} &  \multicolumn{4}{|c|}{All years} & \multicolumn{4}{|c|}{Without 2010} \\             
\hline
Model & a & b & c & d & e & f & g & h & $\Delta$ AICc & Mean ign. & p (IGN) & p (AICc) \\
\hline
M3  & * &   &   &   & * &   &   &   & $-15.30$ & $-0.5864$ & $0.0002$ & $0.0002$ \\
M5  & * &   &   &   & * &   & * &   & $-14.04$ & $-0.4871$ & $0.0002$ & $0.0005$ \\
M10 &   & * &   &   & * &   & * &   & $-12.48$ & $-0.4224$ & $0.0007$ & $0.0024$ \\
M1  & * &   &   &   &   &   &   &   & $-10.89$ & $-0.3920$ & $0.0015$ & $0.0010$ \\
M8  &   & * &   &   & * &   &   &   & $-8.54$  & $-0.2443$ & $0.0081$ & $0.0171$ \\
M2  & * &   &   & * &   &   &   &   & $-8.35$  & $-0.2887$ & $0.0042$ & $0.0042$ \\
\hline
M4  & * &   &   & * &   & * &   &   & $-6.87$  & $-0.0184$ & $0.0132$ & $0.1665$ \\
M6  &   & * &   &   &   &   &   &   & $-4.85$  & $-0.1462$ & $0.0276$ & $0.0305$ \\
M7  &   & * &   & * &   &   &   &   & $-3.94$  & $-0.1189$ & $0.0649$ & $0.0449$ \\
M9  &   & * &   & * &   & * &   &   & $-2.59$  & $-0.1226$ & $0.1123$ & $0.0381$ \\
M14 &   &   & * & * &   & * &   &   & $-1.47$  & $+0.2154$ & $0.2065$ & $0.6255$ \\
M13 &   &   & * &   & * &   &   &   & $-1.21$  & $+0.0342$ & $0.2942$ & $0.6755$ \\
M16 &   &   &   & * & * &   &   & * & $-0.74$  & $-0.0215$ & $0.5430$ & $0.2366$ \\
M11 &   &   & * &   &   &   &   &   & $-0.03$  & $+0.1014$ & $0.2527$ & $0.7715$ \\
M0  &   &   &   &   &   &   &   &   & $0$      & $0$       & -        & - \\
M12 &   &   & * & * &   &   &   &   & $+0.41$  & $+0.1617$ & $0.4324$ & $0.8652$ \\
M15 &   &   & * &   & * &   & * &   & $+1.33$  & $+0.1113$ & $0.4573$ & $0.6875$ \\
\hline
\end{tabular}
\caption{AICc and cross-validated mean ignorance scores expressed relative to the null model.  A star in each column headed by the letters a to h indicates whether the variables listed in section~\ref{section:methods_reindeer} have been included in the model.  Estimated p-values calculated using the single model permutation test are given for each model for each model selection technique.  Models below the horizontal line area included in the second model selection procedure for this data set.}
\label{table:AIC_rel_clim_reindeer}
\end{sidewaystable}

Here, whilst a number of the models are found to be strongly significant, the p-values of those models are typically larger than for the best ibex population models in experiment two. It is therefore prudent to apply the model selection permutation test to assess the probability that the model selection statistics of the `best' model could have occurred by chance.  The results of doing this using the multiple model permutation test are shown in table~\ref{table:sanity_check_reindeer}.  Here, the p-values are small and therefore confidence can be had that the `best' model is indeed informative relative to the null model and did not simply occur by chance. \par

\begin{table}[H]
\centering
\begin{tabular}{|c|c|}
\hline
AICc & Cross-validated mean ignorance \\
\hline
$0.0083$ & $0.0016$ \\
\hline
\end{tabular}
\caption{Estimated p-values from the model selection permutation test for the reindeer example.}
\label{table:sanity_check_reindeer}
\end{table}

The reindeer example is now used to demonstrate a case in which, whilst one or more of the models is found to be significant, the probability that this occurred by chance is found to be high.  Consider a model selection procedure in which the best six models according to the AICc in table~\ref{table:AIC_rel_clim_reindeer} were not used as candidate models and therefore the selection is between the ten remaining models.  The included models are those below the horizontal line in the table.  At least one of the candidate models is significant at the 5 percent level for both model selection techniques.  However, given the number of candidate models, caution is advised.  Applying the model selection permutation test, the p-values shown in table~\ref{table:sanity_check_reindeer_bottom_six} are obtained.  Here, in both cases, the test is insignificant at the 5 percent level and thus there is a high probability that the significance of the individual models simply occurred by chance. \par

\begin{table}[H]
\centering
\begin{tabular}{|c|c|}
\hline
AICc & Cross-validated mean ignorance \\
\hline
$0.0659$ & $0.1902$ \\
\hline
\end{tabular}
\caption{Estimated p-values from the model selection permutation test for the reindeer example when the six best models are removed.}
\label{table:sanity_check_reindeer_bottom_six}
\end{table}

\section{Discussion}
There is a clear and obvious need in ecology for authors to assess the absolute value of the `best' model in a model selection procedure.  Currently, this step is all too often completely absent.  The single model permutation test defined in this paper provides a generalised approach with which to assess the significance of a model.  However, by selecting the `best' model via model selection and proceeding to evaluate its significance, the probability of a type I error can be inflated far beyond the significance level.  This is because the model with the best model selection statistic has already been determined as one that performs well relative to the other models, perhaps by chance. \par

One can imagine that, if each of the models were independent, intuition could be used to assess the impact of multiple testing. Caution would be advised if one out of a total of twenty models were significant at the five percent level, for example.  The Bonferroni correction works on this basis.  Commonly, in model selection in ecology, the same variables are present in multiple models.  Given this dependency, this intuition is lost and therefore more formal methods are required. The model selection permutation test has been proposed for situations such as these.  The test estimates the probability that the `best' model could have occurred by chance, whilst taking the dependency structure between the models into account.  As such, the test gives a clear and intuitive approach to the problem of significance in model selection by assessing the entire model selection procedure. \par

The tests described in this paper can be used to assess whether a set of variables can provide better predictions than the null model in a population modelling procedure.  Although the focus here is on ecology and, in particular, population modelling, the methodology is highly applicable to other fields in which model selection is applied.  For example, in sports forecasting, one may want to determine which combination of factors most impact the probability of scoring a goal or the outcome of a game. \par

Whilst the tests described can help provide confidence that the best candidate variables are more informative than the null model in terms of making predictions, it should be highly stressed that, even if a model can be shown to significantly outperform the null model, it is not necessarily the case that the model is fit for a particular purpose.  Before using the model, further evidence regarding the suitability of the model in a particular setting should be gathered. Nonetheless, the tests described in this paper provide a key step towards rigorous model selection in ecology which, in turn, allows for better modelling and hence a better understanding of the factors that impact animal populations. \par

\appendix

\section{Population Modelling}
The permutation tests described in this paper are demonstrated using two population modelling examples taken from existing papers.  Background methodology relevant to both papers is described here.  Each of the two examples make use of population models.  The Modified Stochastic Gompertz Model is defined by
\begin{equation}
R_{i}=a+b\log(n_{i-1})+\sum_{i=1}^{V} c_{i} V_{i}+\epsilon
\end{equation}
and the Modified Stochastic Ricker model is defined by
\begin{equation}
R_{i}=a+bn_{i-1}+\sum_{i=1}^{V} c_{i} V_{i}+\epsilon
\end{equation}
where $n_{i}$ is the population count in the $ith$ year, $R_{i}=\log(\frac{n_{i+1}}{n_{i}})$ is called the \emph{relative population change}, $V_{i}$ is the $ith$ explanatory variable, and $\epsilon$ is a random draw from a Gaussian distribution with mean zero and variance $\sigma^{2}$.  The two models are very similar and only differ in how the current population is used as an explanatory variable (i.e. which form of so called `density dependence' is considered).  The parameters $a$, $b$ and $c_{1},..,c_{V}$ are to be selected using least-squares estimation.  The Stochastic Gompertz and Ricker Models automatically give probabilistic forecasts of the relative population change in the form of of a Gaussian distribution $N(\hat{R_{i}},\sigma^{2})$. The forecast distribution of the relative population change can be used to estimate a forecast distribution of the actual population.  In this paper, where applicable (for the reindeer case), this is done using Monte-Carlo simulation with $10,000$ samples.  \par

A `null' model distribution naturally arises from the Modified Stochastic Ricker or Gompertz Model with all parameters except for the intercept and the variance set to zero.  The null model therefore takes the form $R_{i} \sim N(a,\sigma^{2})$ where $a$ and $\sigma$ are parameters to be selected. \par

\bibliographystyle{agsm}
\bibliography{bibliography}

\end{document}